\documentclass{emulateapj}
\usepackage{apjfonts}

\newcommand{\kms}{\mbox{km s$^{-1}$}}
\newcommand{\hgyr}{\mbox{h$^{-1}$Gyr }}
\newcommand{\hmyr}{\mbox{h$^{-1}$Myr }}
\newcommand{\hkpc}{\mbox{h$^{-1}$kpc }}
\newcommand{\hpc}{\mbox{h$^{-1}$pc }}
\newcommand{\hmsun}{\mbox{h$^{-1}$M$_{\sun}$ }}

\newcommand{\lir}{\mbox{L$_{\rm IR}$ }}

\newcommand{\cmtwo}{\mbox{cm$^{-2}$ }}

\shorttitle{Molecular Outflows in Galaxy Mergers with Embedded AGN}
\shortauthors{Narayanan et al.}

\begin{document}
\title{Molecular Outflows in Galaxy Merger Simulations with Embedded
AGN} \author{Desika Narayanan\altaffilmark{1,4}, Thomas
J. Cox\altaffilmark{2}, Brant Robertson\altaffilmark{2}, Romeel
Dav\'{e}\altaffilmark{1}, Tiziana Di Matteo\altaffilmark{3}, Lars
Hernquist\altaffilmark{2}, Philip Hopkins\altaffilmark{2}, Craig
Kulesa\altaffilmark{1}, Christopher K. Walker\altaffilmark{1}}

\altaffiltext{1}{Steward Observatory,
University of Arizona, 933 N Cherry Ave, Tucson, Az, 85721, USA}
\altaffiltext{2}{Harvard-Smithsonian Center for Astrophysics, 
60 Garden Street, Cambridge, MA 02138, USA}
\altaffiltext{3}{Carnegie Mellon University, 
Department of Physics, 5000 Forbes Ave., Pittsburgh, PA 15213}
\altaffiltext{4} {dnarayanan@as.arizona.edu}

\slugcomment{Accepted to The Astrophysical Journal Letters}

\begin{abstract}

We study the effects of feedback from active galactic nuclei (AGN) on
emission from molecular gas in galaxy mergers by combining
hydrodynamic simulations which include black holes with a
three-dimensional, non-local thermodynamic equilibrium (LTE) radiative
transfer code.  We find that molecular clouds entrained in AGN winds
produce an extended CO morphology with significant off-nuclear
emission, which may be detectable via contour mapping.  Furthermore,
kinematic signatures of these molecular outflows are visible in
emission line profiles when the outflow has a large line of sight
velocity.  Our results can help interpret current and upcoming
observations of luminous infrared galaxies, as well as provide a
detailed test of subresolution prescriptions for supermassive black
hole growth in galaxy-scale hydrodynamic simulations.

\end{abstract}


\section{Introduction}

The physical processes giving rise to the birth and sustained fueling
of massive starbursts and active galactic nuclei (AGN) have been of
interest since their discovery.  It is generally agreed that AGN are
powered by accretion of gas onto supermassive black holes (BHs) in the
centers of galaxies (e.g. Lynden-Bell, 1969), but the fueling
mechanism is less clear.  Hydrodynamic simulations have shown that
mergers can produce strong, galaxy-scale inflows owing to
gravitational torques (Barnes \& Hernquist 1991, 1996), triggering
starbursts (Mihos \& Hernquist, 1996).  This suggests a circumstantial
link between starbursts and AGN activity.

Observational evidence connecting the evolution of starbursts into AGN
through mergers is compelling. Some ultraluminous infrared galaxies
(ULIRGs, \lir $\geq$10$^{12} L_\sun$) exhibit spectral energy
distributions (SEDs) characteristic of classical starbursts, whereas
others have SEDs more closely resembling optical quasars (e.g. Farrah
et al. 2003).  In seminal papers, Soifer et al. (1987), and Sanders et
al. (1988a,b) used optical, near infrared and millimeter-wave
observations to advance a scenario in which starburst dominated ULIRGs
served as antecedents of AGN.  The specifics of a starburst-AGN
connection, however, remain under heavy debate throughout the
literature (e.g. Sanders \& Mirabel, 1996).

Recent numerical models by Di Matteo, Springel \& Hernquist (2005),
Hopkins et al. (2005a-d), and Springel, Di Matteo \& Hernquist (2005a)
have provided a theoretical foundation for the link between starbursts
and AGN.  In particular, by modeling the growth of (and feedback from)
central black holes, they showed that gas-rich galaxy mergers are a
viable candidate to serve as a precursor to the formation of quasars.
Their simulations also show that feedback from accreting black holes
in galaxies are relevant to a wide range of phenomena associated with
the evolution of galaxies in mergers including characteristic X-ray
emission patterns (Cox et al. 2006), observed quasar luminosity
functions and lifetimes (Hopkins et al. 2005c), the $M_{\rm
BH}-\sigma$ relation (Di Matteo et al 2005; Robertson et al. 2006a),
the fundamental plane of ellipticals (Robertson et al 2006b), and the
bimodal galaxy color distribution (Springel, Di Matteo \& Hernquist,
2005b; Hopkins et al. 2005e).

There has been a longstanding interest in better understanding the
nature of the molecular interstellar medium (ISM) in mergers, as the
molecular gas serves as fuel for the induced starburst activity and
possibly for accreting BH(s).  High resolution observations have
identified massive concentrations of molecular gas in the nuclear
regions of ULIRGs (Bryant \& Scoville, 1999), as well as high
excitation molecular gas in regions of massive starbursts (Iono et
al. 2004; Wang et al. 2004).

Large surveys at submillimeter (sub-mm) and millimeter wavelengths
have shown that mergers in the local Universe emit copious molecular
line radiation both from diffuse molecular gas (e.g. Sanders, Scoville
\& Soifer, 1991, among others), and dense cloud cores (e.g. Gao \&
Solomon, 2004; Narayanan et al. 2005).  Other work demonstrates that
high-$z$ infrared luminous and sub-mm selected sources also contain
significant amounts of molecular gas (Greve et al. 2005; Tacconi et
al. 2006).  Moreover, as evidenced by IR and X-ray studies, a large
fraction of these galaxies at high-$z$ contain AGN (e.g. Alexander et
al. 2005, Polletta et a. 2006).  However, despite the wealth of data on
molecular emission in interacting galaxies, little is known about the
impact of embedded AGN on this radiation.

Here, we describe preliminary attempts to quantify the observable
effects of AGN feedback on molecular line emission from major galaxy
mergers.  We use hydrodynamic simulations of mergers with and without
AGN, combined with a new 3D non-LTE radiative transfer code (Narayanan
et al. 2006a,b) to model CO emission. In this {\it Letter}, we describe
results in which we find distinct signatures of AGN feedback on cold
gas, and discuss some observational results that may be understood in
this context.

\section{Numerical Simulations}

Our hydrodynamic simulations were performed using the
$N$-body/smoothed particle hydrodynamics (SPH) code, GADGET-2
(Springel, 2005). This code uses a fully conservative formulation of
SPH (Springel \& Hernquist 2002), and accounts for radiative cooling
of the gas (Dav\'e et al. 1999), a multi-phase description of the ISM
which includes cold clouds in pressure equilibrium with hot, diffuse
gas (e.g. McKee \& Ostriker, 1977), and a prescription for star
formation constrained by the Schmidt/Kennicutt laws (Kennicutt, 1998,
Schmidt, 1959; see Springel \& Hernquist, 2003). The black hole(s) in
the simulation are realized through sink particles which accrete gas
from the surrounding ISM such that 0.5\% of the accreted mass energy
onto the central black hole(s) is reinjected into the ISM as thermal
energy (Di Matteo et al. 2005, Springel et
al. 2005a,b).

The progenitor disk galaxies used in this work are similar to the
Milky Way, but with a higher gas fraction (50\%).  Thus, they are
likely representative of high-$z$ disk galaxies, and possible
progenitors of present epoch ULIRGs. The methodology for constructing
the model galaxies is given in Springel et al.  (2005a).  The
progenitors utilized a softened equation of state (EOS) with softening
parameter $q_{\rm EOS}$=0.25 (Springel et al. 2005a) such that the
mass-weighted ISM temperature is $\sim$10$^{4.5}$ K.  The galaxies had
dark matter halos initialized to follow a Hernquist (1990) profile,
and circular velocity $V_{200}$=160 \kms.  The virial properties of
the halos are scaled to be appropriate for $z=2$ (e.g. Robertson et
al. 2006a), and follow the prescription given by Mo et al.  (1998) for
cosmological models of disk galaxies. The galaxies were set on a
parabolic orbit with the orientation of the spin axis of each disk
specified by the standard spherical coordinates, $\theta$ and $\phi$
(with $\theta_1$=30\degr, $\phi_1$=60\degr, $\theta_2$=-30\degr,
$\phi_2$=45\degr), and were initially separated by 140 kpc . We
utilize 120,000 dark matter particles, and 160,000 total disk
particles, 50\% of which represent gas, the rest serving as
collisionless star particles. The gas, star, disk and dark matter
particle masses were 3.9$\times$10$^5$, 1.95$\times$10$^5$,
5.9$\times$10$^5$ and 7.6$\times$10$^6$ \hmsun each, respectively. The
gravitational softening lengths were 100 \hpc for baryons, and 200
\hpc for dark matter particles.  We have performed simulations with
and without black holes. For the simulations with BHs, the initial
mass of the BH particle in each progenitor 10$^5$ \hmsun, and the peak
accretion rate was $\sim$0.5 M$_\sun$yr$^{-1}$. The final mass of the
remnant's BH is $\sim$5$\times$10$^7$ \hmsun.


In order to estimate molecular line emission, we have developed a
three dimensional non-local thermodynamic equilibrium (LTE) radiative
transfer code based on an improved version of the Bernes (1979)
algorithm. Our improvements (described more fully in Narayanan et
al. 2006a,b) focus on including a subgrid model for giant molecular
clouds (GMCs) in order to more accurately model the strongly
density-dependent collisional excitation rates within our
$\sim$10$^2$pc grid cells.  We model GMCs as singular isothermal
spheres (SISs) with power law index 2 (Walker, Adams \& Lada, 1990),
and assume half the cold gas mass in each cell is bound in GMCs.  Our
results for higher lying molecular levels and high dipole moment
molecules are particularly improved using this subgrid approach.

The emergent spectrum is built by integrating the equation of
radiative transfer through numerous lines of sight. The source
functions in each grid cell are determined by the densities of
molecules at levels $u$ and $l$ for a given transition $u \rightarrow
l$. These level populations, $n_u, n_l$ are dependent on the incident
radiation field from other clouds; we thus guess at a solution,
calculate the mean intensity field in a Monte Carlo manner, determine
the updated level populations through the rate equations, and repeat
the process until the populations have converged. In between each
iteration, the code goes into a sub-resolution process in which it
decomposes each cloud into a SIS and determines the level populations
in each sub-resolution element. The non-LTE aspect of this treatment
is particularly important as the assumption of LTE breaks down when
considering the propagation of radiation through media with densities
lower than the transition's critical density. For the radiative
transfer calculations, we typically emitted $\sim$1$\times$10$^7$
model photons per iteration, and consider the 2.73 K microwave
background as the boundary condition.

\section{Results}
\subsection{Intensity Contour Maps}

\begin{figure}
\includegraphics[scale=.9]{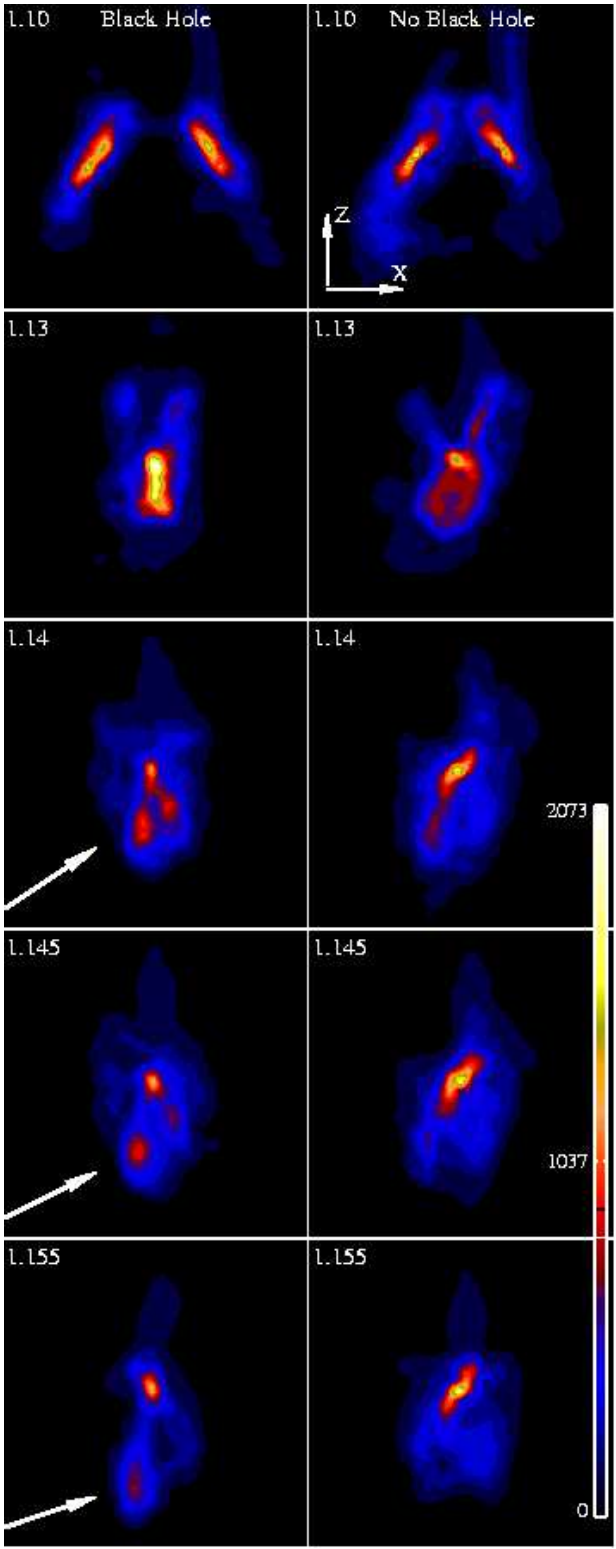}
\caption{CO (J=1-0) emission intensity contours for an equal mass
galaxy merger. The left column is a time sequence for the model with
BHs, and the right column without BHs. In the model with BHs, after
the galaxies merge, massive amounts of gas are driven into the nuclear
regions and accrete onto the supermassive black hole. The subsequent
AGN feedback energy can blow large blobs of molecular gas out from the
nuclear regions (see arrows). These features are not seen in the model
without BHs. The time stamp of the image is in the top left of each
panel, and is in units of \hgyr. The color contours are in units of
K-\kms (velocity-integrated Rayleigh-Jeans temperature), and the scale
is on the right side of the plot. Each panel is 12 \hkpc on a
side. The images are made at 1/4 \hkpc spatial resolution. The
coordinate system for use with Figure~\ref{figure:line} is in the top
right panel.
\label{figure:comerger}}
\end{figure}

In Figure~\ref{figure:comerger}, we show a series of snapshots in CO
(J=1-0) intensity contour maps from the two merger simulations. The plot
spans 45 \hmyr, and begins when the progenitors are approaching final
coalescence. The black hole accretion rate nears its peak as the black
holes merge at T$\sim$1.15 \hgyr, and thus the feedback energy input
from the AGN is near its maximum in the model with BHs.

Beginning from T$\approx$1.13 \hgyr onward, the CO morphology of the
galaxy in the BH model undergoes dramatic changes owing to feedback
from the buried AGN. Massive blobs of cold molecular gas entrained in
the wind are visible through the CO (J=1-0) tracer. Indeed, while the
AGN wind may in detail evaporate cold clouds via thermal
conduction, enough cold gas survives to be visible through molecular
emission.  Moreover, the clouds entrained in these outflows remain
cold and dense enough to continue forming stars. The dense cores in
these clouds emit at CO transitions with relatively high critical
densities; consequently, the outflows are visible against the
background at transitions as high as CO (J=6-5). In order to produce
the observed emission comparable to that of the nucleus, the outflows
must have large column densities. We find columns ranging from
5$\times$10$^{22}$\cmtwo $\lesssim$ $N$(H$_2$) $\lesssim$
1$\times$10$^{23}$\cmtwo through the outflow in
Figure~\ref{figure:comerger}, depending on the viewing angle
(although, we did not include a UV background in our models, and thus
the true column may be less).  As the outflowing gas leaves the nuclear
region at velocities of $\sim$200-300 \kms, it becomes more diffuse,
resulting in weaker CO emission. We find that the existence of
molecular outflows is not unique to this particular model, and is seen
in other merger simulations which include BHs.

In the model without BHs (right column, Figure~\ref{figure:comerger}),
the molecular outflows on the $\sim$kpc scale observed in the BH model
are not seen, highlighting the effects of AGN feedback in expelling
loosely bound circumnuclear molecular gas. We note, however, that
supernovae-driven winds are not incorporated in these models. It is
known, through absorption line spectroscopy, that such winds in
starbursts can induce outflows of comparable speeds (e.g. Heckman et
al. 2000; Martin, 2005; Rupke et al. 2005). While large
columns of outflowing molecular gas have not been imaged in many of
these systems, there are notable exceptions such as the classic
starburst M82 (Walter, Wei\ss \ \& Scoville, 2002). We will explore
the effects of supernovae winds on the molecular gas in due course.

\subsection{Line Profiles}

\begin{figure*}
\plotone{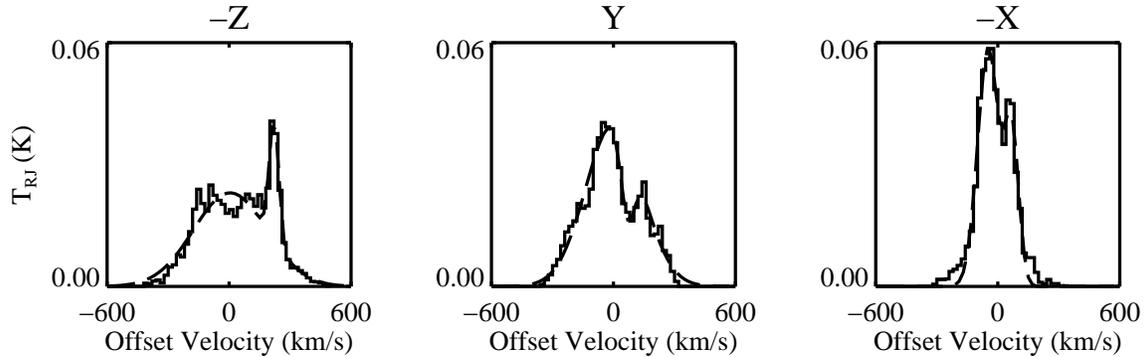}
\caption{Synthetic CO (J=1-0) (rest-frame) emission line profile taken
at T=1.155 \hgyr in the merger run with black holes over 3 orthogonal
viewing angles. The direction of each line of sight is above the
panels, and corresponds with the coordinate system in the top right
panel of Figure~\ref{figure:comerger}. The high velocity peak in the
leftmost panel owes to the outflowing gas viewed with a large line
of sight velocity component. In the $\hat{y}$ direction, the outflow
has a smaller LOS velocity, and the emission owing to the outflow blends
more into the main emission. In the -$\hat{x}$ viewing angle, outflows
have a negligible effect on emission line.  The best fit to the model
spectra are overplotted with dashed lines.  The spectra is taken such
that the object is at $z=$2, and convolved with a 30$\arcsec$ circular
Gaussian beam.}
\label{figure:line}
\end{figure*}

The AGN induced outflows also leave their imprint on spectral line
profiles. In Figure~\ref{figure:line}, we have calculated the spectral
line emission from T= 1.155 \hgyr in the model with black holes (lower
left panel in Figure~\ref{figure:comerger}). The spectrum is generated
along three orthogonal lines of sight. In order to simulate an
unresolved observation, we set the merger at $z$=2
($\Omega_\Lambda$=0.7, $\Omega_M$=0.3, $h$=0.75), and convolved the
emission from the 12\hkpc image with a circular 30$\arcsec$ ($\sim$235
kpc) Gaussian beam.

Typically, in the BH model, once the galaxies have coalesced (when the
BHs of the progenitors are indistinguishable in our simulations), the
emission from the unresolved object is characteristic of a single
Gaussian, centered at the systemic velocity of the galaxy. However,
when viewing outflows with a strong line of sight (LOS) velocity
component, a secondary peak appears superposed on the Gaussian
emission line from the galaxy (e.g. left panel,
Figure~\ref{figure:line}); this peak is the emission from the outflow,
and is redshifted or blueshifted from line center at the LOS velocity
of the outflow.  While we have presented the CO (J=1-0) emission,
these line profiles are similar through CO (J=6-5).

The emission peak corresponding to the outflow appears at the greatest
offset velocity with respect to the systemic velocity of the galaxy
when the outflow is moving mostly along the LOS (e.g. the -$\hat{z}$
observation, Figure~\ref{figure:line}).  If the observation is tilted
such that a smaller component of the outflow velocity is along the
LOS, the emission peak corresponding to the outflow will move closer
to, and eventually will merge with the broad emission peak of the
galaxy ($\hat{y}$ observation and -$\hat{x}$ observation,
Figure~\ref{figure:line}, respectively). The peak temperature of the
outflow emission may also decrease when the observation is tilted as
the observer looks through less column. A given outflow along a
particular line of sight is typically visible via its line profile for
an average of $\sim$10 \hmyr before the column density through the
outflow drops such that its emission is no longer detectable against
the broader Gaussian emission from the central region. We estimate the
``outflow'' profiles are visible $\sim$25\% of the time in our
simulations (which span 200 \hmyr), averaged over many viewing angles.

Double-peaked profiles have been observed in mergers which do not
necessarily correspond to outflows: for example, observations of high
density gas in the prototypical ULIRG, Arp 220, have evidenced a
symmetric double-peak profile where each peak corresponds to the
starburst regions of nuclei of the progenitor galaxies (Taniguchi \&
Shioya, 1998, Sakamoto et al. 1999, Narayanan et al. 2005).  Similar
profiles also occur in systems in which there is significant rotation,
i.e. a disk galaxy or rotating nuclear ring. However, some differences
exist between the double peaks originating in progenitor galaxies or
rotating systems, and those caused by outflows. The double peaks
characteristic of the former two cases are typically both broad, and
symmetric about the systemic velocity of the galaxy. Conversely, the
peak arising from the outflow is typically much narrower than the
broad emission profile of the galaxy, due to its small velocity
dispersion along the LOS. That said, the component of the profile from
the outflow can be quite bright, owing to large H$_2$ column densities
through the outflowing material.

The characteristic outflow profile consisting of a broad Gaussian with
a narrow line superposed is also degenerate with that of high velocity
gas falling in toward the nucleus. The infalling clouds tend to have a
higher velocity dispersion than clouds entrained in the AGN wind by,
on average, $\sim$ 30\%, resulting in broader CO lines. It is unclear,
though, from our simulations whether there is a significant enough
difference between the velocity dispersion or column density in
outflowing and infalling gas to determine the direction of flow from
CO observations alone. There may be observational tests at other
wavelengths to help break this degeneracy, however.  In our
simulations, infalling gas primarily leaves its imprint on the line
profile just as the major merger has occurred, prior to the major AGN
feedback phase. By serving as an indicator for AGN activity, X-ray
observations may help distinguish the origin of the ``outflow'' line
profile.  For example, using similar simulations to those explored in
this study, Cox et al. (2006) have shown that the X-ray luminosity
from diffuse gas in galaxy mergers peaks during the phase of heavy
black hole accretion, and that thermal energy input from AGN feedback
can produce X-rays consistent with observations of ULIRGs with known
embedded AGN (e.g. UGC 5101, Imanishi et al. 2003). Elevated X-ray
emission, both from diffuse gas, as well as hard X-rays from the
central black holes may be indicative of a buried AGN. Correlations
between hard X-ray flux and CO line profiles will be discussed in
greater detail in Narayanan et al. (2006b).

At z$\approx$2, where the infrared and submillimeter background are
dominated by LIRGs and ULIRGs (Smail, Ivison \& Blain, 1997), and the
quasar density is near its peak (e.g. Schneider et al. 2005), large
molecular line surveys may prove fruitful in investigating the
existence of a correspondence between double-peaked line profiles and
quasar activity. Indeed, large fractions of sub-mm galaxies at high
redshift show double-peaked line profiles ($\sim$ 50\%, Greve et
al. 2005, Tacconi et al. 2006), some of which appear to have line
profiles similar to those presented in the leftmost and middle panels
of Figure~\ref{figure:line}; similarly, large fractions ($\sim$ 75\%)
of sub-mm galaxies show AGN activity as evidenced by IR and X-ray studies
(Alexander et al. 2005, Polletta et al. 2006).

\section{Summary and Conclusions}
\label{sec:disc}

We have discussed two features of CO emission from cold molecular gas
entrained in winds resulting from AGN feedback in galaxy mergers: 1.)
an extended CO morphology when the outflow is largely in the plane of
the observation, and 2.) kinematic features in the emission line
profile when the outflow has a significant line of sight velocity
component. There may, of course, be hybrid cases as well, in which
features of both signatures of AGN feedback are detectable.

Observations of signatures such as these can help to interpret current
observations of ULIRGs both at low and high-$z$. Emission maps and
line profiles similar to those presented in this work may have already
been observed in local mergers (e.g. NGC 985, Appleton et al. 2001),
as well as in $z\approx 2$ sub-mm selected sources (Greve et al. 2005,
Tacconi et al. 2006). Upcoming surveys with high resolution
interferometers have the potential of constraining models of black
hole growth and associated AGN feedback in galaxy mergers.

\acknowledgements D.N. acknowledges financial support from an NSF
Graduate Research Fellowship, and thanks Brandon Kelly and Casey
Papovich for helpful conversations. The calculations were performed on
Grendel, a Steward Observatory Beowulf system.

\clearpage

\clearpage

\clearpage





\end{document}